%% file: main.tex
\documentclass[conference]{IEEEtran}

\IEEEoverridecommandlockouts
\usepackage{cite}
\usepackage{amsmath,amssymb,amsfonts}
\usepackage{algorithmic}
\usepackage{graphicx}
\usepackage{textcomp}
\usepackage{xcolor}
\usepackage{balance}

\usepackage{booktabs}
\usepackage{xcolor}
\usepackage{subfig}
\usepackage{graphicx}

\usepackage{amsmath}
\usepackage{mathtools}
\usepackage{amsthm}
\usepackage{amssymb}

\usepackage{array}

\usepackage{siunitx}

\usepackage{url}

\usepackage{fancyhdr}

\newcommand{\niparagraph}[1]{\vspace{1pt}\noindent\textbf{#1}}

\def\BibTeX{{\rm B\kern-.05em{\sc i\kern-.025em b}\kern-.08em
    T\kern-.1667em\lower.7ex\hbox{E}\kern-.125emX}}

\definecolor{brickred}{rgb}{0.8, 0.25, 0.33}
\colorlet{shadecolor}{gray!20}

\begin{document}

\title{\Huge{Characterizing Compute-Communication Overlap in GPU-Accelerated Distributed Deep Learning: Performance and Power Implications}}

%\author{\IEEEauthorblockN{Anonymous Authors}}

\author{
Seonho Lee$^{\dagger}$,
Jihwan Oh$^{\dagger}$$^{\ddagger}$,
Junkyum Kim$^{\dagger}$,
Seokjin Go$^{\dagger}$,
Jongse Park$^{\ddagger}$,
Divya Mahajan$^{\dagger}$ \\
\\
\normalsize{
$^{\dagger}$Georgia Institute of Technology \quad
$^{\ddagger}$KAIST \quad
\vspace{0.3em}}
\\
\normalsize{
{seonho.lee@gatech.edu},
{jihwan.oh1@gatech.edu},
{jun-kyum.kim@gatech.edu},
{seokjin.go@gatech.edu},} \\
\normalsize{
{jspark@casys.kaist.ac.kr},
{divya.mahajan@gatech.edu}}
}

 \maketitle

% enable page numbers
\thispagestyle{plain}
\pagestyle{plain}

\input{body/abstract}

\input{body/introduction}
\input{body/background}
\input{body/methodology}
\input{body/evaluation}

\input{body/related}
\input{body/conclusion}

\input{body/ack}

\balance

\bibliographystyle{IEEEtran}
\bibliography{references}

\end{document}

%% file: body/abstract.tex
\begin{abstract}
This paper provides an in-depth characterization of GPU-accelerated systems, to understand the interplay between overlapping computation and communication which is commonly employed in distributed training settings. 
Due to the large size of models, distributing them across multiple devices is required.
Overlapping strategies, which enable concurrent computation and communication, are critical for mitigating communication bottlenecks and maximizing GPU utilization.
However, the current consensus is that we should always and aggressively overlap compute and communication to mitigate the overhead of distribution~\cite{deepspeed, pipedream, zero}.
By systematically evaluating state-of-the-art GPUs, including NVIDIA's H100 and A100, and AMD's MI250 and MI210, this study investigates the impact of hardware features such as numeric precision, specialized cores, and power capping on distributed training workloads.

Comprehensive experiments and studies showcase the effects of overlapping strategies on performance and power consumption across varying scenarios.
We observe that overlapping computation and communication can result in an average computational slowdown of 18.9\%, with a maximum of 40.0\% slowdown.
This slowdown is in comparison to the scenario when no communication was happening with the compute. We consider this an ideal execution scenario, where the communication in parallel has not impact on the compute time. 
However, performing computation and communication sequentially is, on average, 10.2\% slower than overlapped execution, with a maximum slowdown of 26.6\%.
We perform these experiments across state-of-the-art GPT and LLaMA models, across 4 A100s, H100s, MI250, and MI210. 
We further observe, while specialized datapath and optimized numeric precision mitigate certain slowdowns, overlapping execution can lead to resource contention and also increase power consumption under specific configurations. 
The analysis also uncovers trade-offs introduced by power and frequency capping, emphasizing the importance of balanced strategies to optimize energy efficiency and training throughput. 

\end{abstract}

%% file: body/introduction.tex
\section{Introduction}

Deep learning has revolutionized fields such as natural language processing, computer vision, and scientific computing, driven by increasingly large models like GPT and LLaMA~\cite{gpt2, gpt3, llama, llama3}. These models, often exceeding billions of parameters, require immense computational power, often provided by GPU-accelerated systems.
To support the execution of large models, we require distributed execution to cater to their growing memory requirements.
In distributed training, GPUs collaboratively compute and exchange information to synchronize model parameters and send/receive model activations and gradients~\cite{pytorch}. 
As model sizes and the number of devices hosting the model grows, the cost of communication increases proportionally, leading to idle GPU time caused by data dependencies during communication~\cite{deepspeed, deepspeedinference}. 
Thus, latest distributed training frameworks and optimizers such as ZeRO and Pipedream attempt to mitigate these inefficiencies by overlapping communication with computation~\cite{zero,pipedream,pipedream2bw}. 
Other works in recommender models also enable such optimizations~\cite{fae, hotline}.
Communication tasks without direct data dependencies on subsequent computations are overlapped, effectively hiding communication latency and reducing idle GPU time.

\begin{figure}[t]
\centering
\subfloat[Fully-Sharded Data Parallelism]{\includegraphics[width=0.8\linewidth]{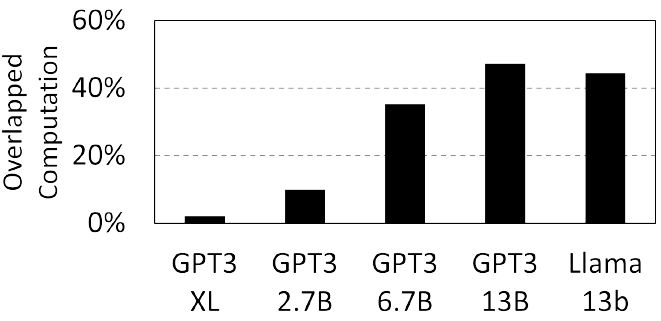}}\\
\subfloat[Pipeline Parallelism]{\includegraphics[width=0.7\linewidth]{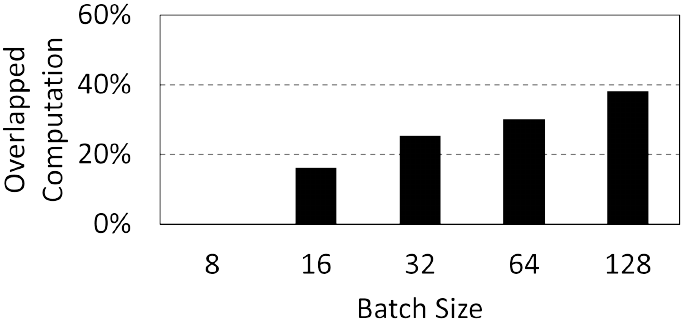}}
\caption{Amount of overlapping computation and communication across varying model sizes and batch sizes. (a) H100x8 system with Fully-Sharded Data Parallelism; (b) A100~$\times$~4 system with Pipeline Parallelism and GPT3 2.7B.}
\label{fig:motivation}
\vspace{-2ex}
\end{figure}

Figure \ref{fig:motivation} illustrates this trend. As model size or batch size increases, the proportion of computation overlapped with communication also increases, attempting to hide the communication overhead. However, as overlap grows, it introduces complexities that affect both performance and power efficiency, warranting a closer examination of these interactions.
Despite these advancements, prior works often overlook the interplay between overlapping computation and communication. These frameworks frequently assume constant computation and communication latencies, neglecting their dynamic interaction. Our work demonstrates that overlapping significantly influences both performance and power consumption in ML training. This observation is particularly critical as scaling model training introduces aggressive optimizations, such as those proposed recently~\cite{domino}, which heavily rely on such overlapping techniques.

\begin{figure*}[t!]
\centering
\subfloat[GPU architecture highlighting key components~\cite{nvidia_gpus, amd_gpus}.\label{fig:gpu_arch}]{\includegraphics[width=0.25\textwidth]{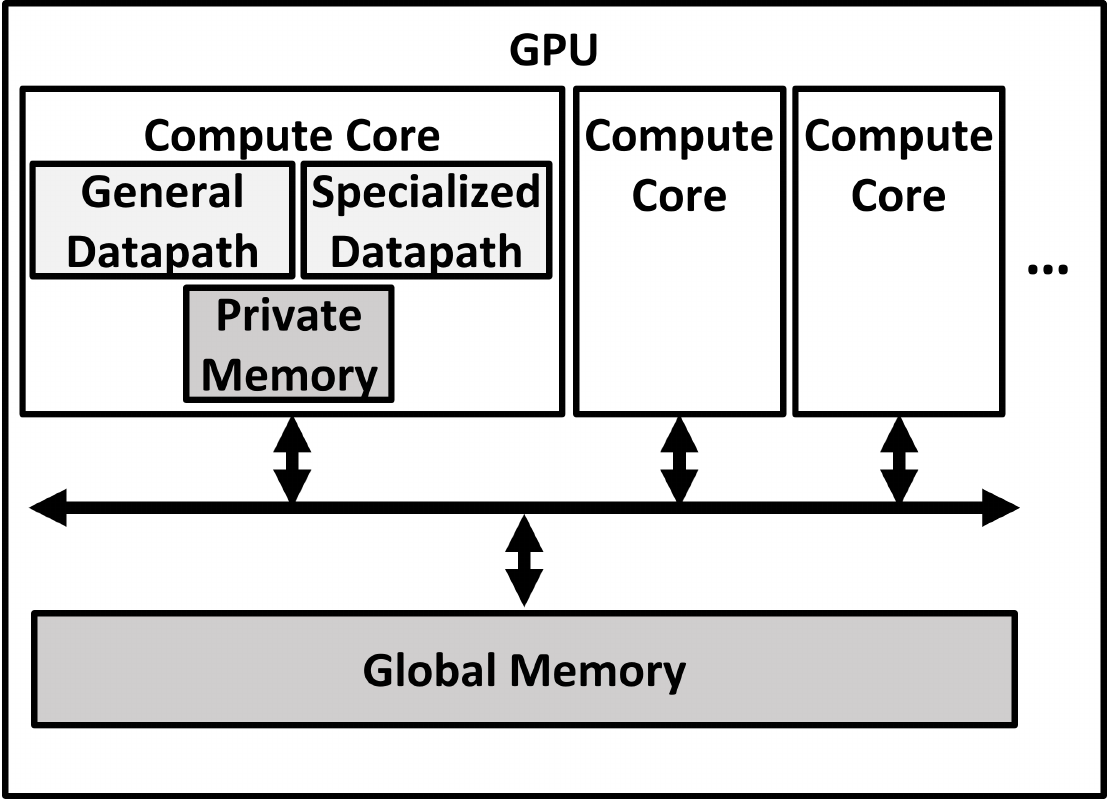}}
\hspace{10ex}
\subfloat[Overview of server interconnects used for distributed training.\label{fig:gpu_server}]{\includegraphics[width=0.55\textwidth]{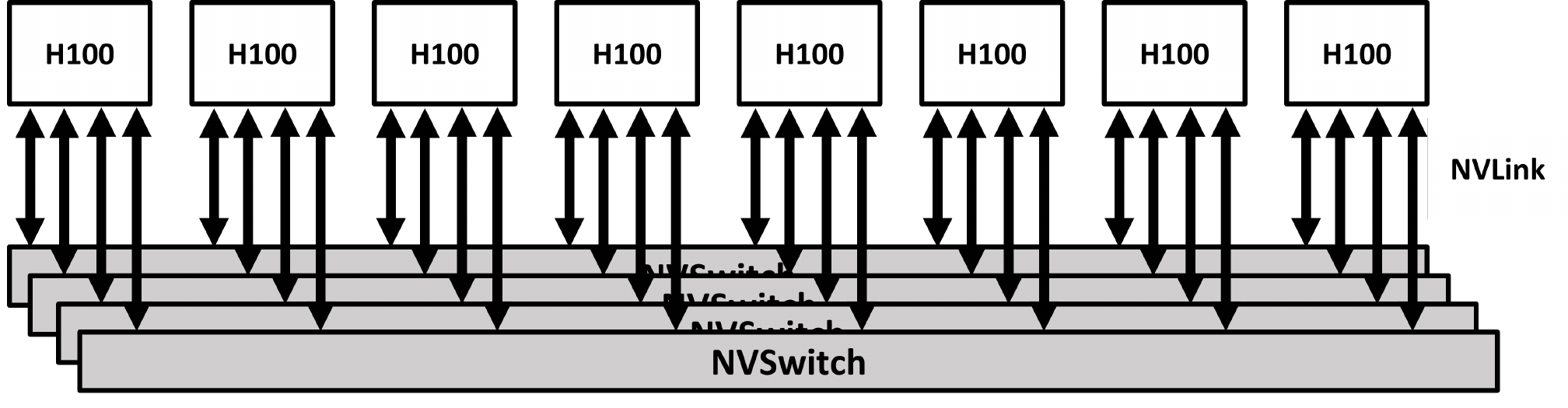}}
\caption{GPU server architecture, illustrating (a) the internal GPU design and (b) the interconnect setup in H100 DGX Box~\cite{nvidia-dgx}.}
\label{fig:gpu}
\vspace{-2ex}
\end{figure*}

Motivated by these challenges, this study provides a comprehensive characterization of GPU-accelerated distributed training setups. We evaluate state-of-the-art hardware, including NVIDIA’s H100 and A100 GPUs and AMD’s MI250 and MI210 GPUs~\cite{h100, a100, mi250, mi210}, to uncover critical interactions between hardware features—such as power and frequency limits, numeric precision, and specialized cores—and distributed training configurations like data parallelism, pipeline parallelism, batch sizes, and model architectures.
Our methodology measures the performance and power impacts of overlapping computation and communication on compute performance in distributed setups, evaluating hardware features like power/frequency capping, specialized datapaths, and different numeric precisions.
The insights presented in this paper aim to advance the performance and power efficiency of GPU-accelerated distributed deep learning systems, enabling faster and more scalable training of large-scale models. These findings serve as a foundation for future research on addressing the challenges of scaling deep learning workloads in an era of ever-growing AI demands.

Our experiments on GPT and LLaMA models using A100s, H100s, MI250s, and MI210s show that overlapping computation and communication leads to an average compute slowdown of 18.9\%, with a maximum of 40.0\%, compared to an ideal execution scenario where compute and communication have no effect on each other. 
Sequential execution, in contrast, is on average 10.2\% slower than overlapped execution, with a maximum slowdown of 26.6\%.
This highlights a gap between the ideal execution time and the overlapped execution time. While overlapped execution outperforms sequential execution in terms of performance, it experiences resource contention leading to slowdown compared to the ideal execution time where compute and communication do not interfere with each other.
Additionally, while overlapping strategies effectively hide communication overhead, they can significantly increase power consumption under specific configurations. 
Under strict power caps, overlapping execution experiences severe slowdowns, with execution time increasing by up to 107\%, indicating that power constraints amplify resource contention effects.
These findings showcase the importance of balancing performance and resources such as energy efficiency to optimize distributed training systems effectively.

%% file: body/background.tex
\section{Background}

\subsection{GPU Architecture}

Modern deep learning training heavily relies on the Graphics Processing Unit (GPU) for its advancement, which offers the parallel computational power necessary for handling large-scale data operations. Modern GPUs from various vendors, including Nvidia and AMD~\cite{h100,a100,l4,t4,mi250x,mi250,mi210,mi100,nvidia_gpus,amd_gpus}, have evolved to support the demanding workloads of deep learning. These GPUs feature high computational throughput, extensive memory bandwidth, and specialized units designed to accelerate matrix operations and other AI workloads~\cite{h100_whitepaper,cdna2}. Advances such as increased core counts, higher memory capacities, and enhanced interconnect technologies have been instrumental in enabling scalable distributed training setups across multiple devices and nodes.

\begin{figure*}[t!]
\centering
\includegraphics[width=1.0\textwidth]{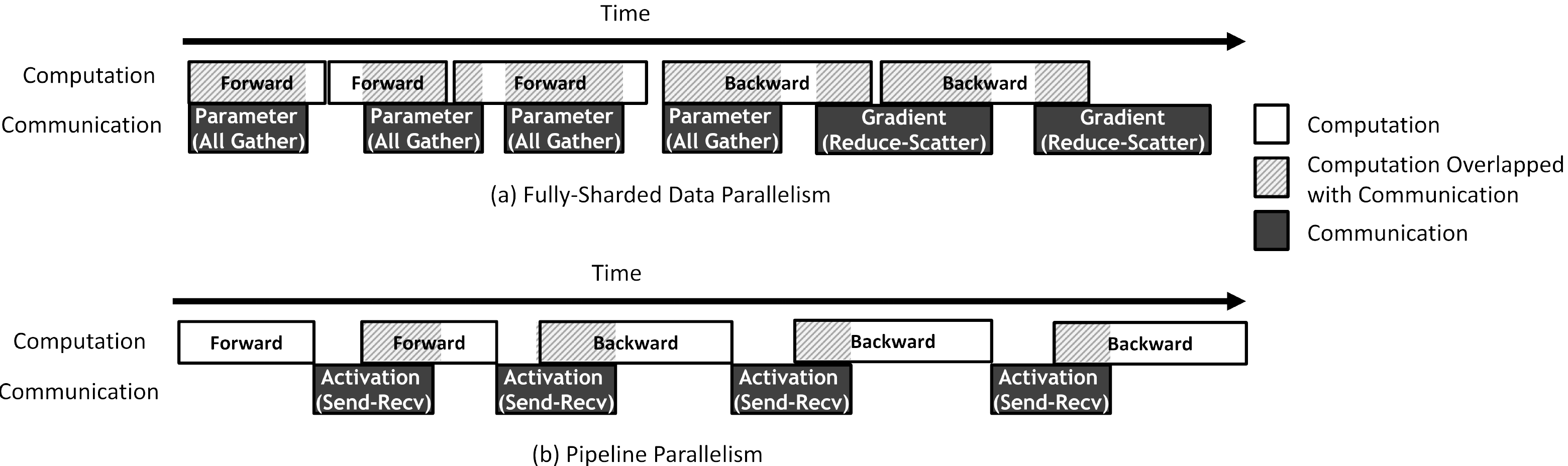}
\caption{Illustration of distributed training techniques: (a) Fully Sharded Data Parallelism (FSDP) and (b) Pipeline Parallelism. The periods of overlapping computation and communication are highlighted.}
\label{fig:distributed_training}
\vspace{-2ex}
\end{figure*}

GPU servers, such as the NVIDIA DGX H100 system~\cite{h100dgx} shown in Figure~\ref{fig:gpu_server}, are designed to integrate multiple GPUs within a single node for high-performance workloads. These servers feature high-bandwidth interconnects like NVLink and NVSwitch, enabling low-latency communication between GPUs~\cite{nvswitch,Li_2020}. The design of the GPU architecture focuses on minimizing bottlenecks in memory access and inter-GPU communication to maximize throughput and efficiency during training. Similarly, AMD Instinct servers utilize Infinity Fabric to connect GPUs, achieving high-bandwidth communication and scalability across larger clusters~\cite{amd-miserver,amdmultigpu}. The combination of these interconnect technologies and optimized server designs ensures efficient resource utilization in distributed deep learning frameworks.
In this work, we consider systems similar to the aforementioned ones, where GPUs are interconnected using high-bandwidth links and switches. Specifically, we include DGX H100 and A100 systems interconnected with NVLink and NVSwitch, as well as MI250 and MI210 systems interconnected with Infinity Fabric.

\subsection{Deep Learning Frameworks and Distributed Training}

Frameworks such as DeepSpeed and Horovod facilitate distributed training, providing tools for distributing data, managing parameters, and synchronizing gradients across multiple devices~\cite{deepspeed, horovod}. 
In training, tasks are partitioned across multiple GPUs in a distributed fashion, often involving two types of parallelism: data parallelism and model parallelism~\cite{dataparallel, megatronlm}. 
Data parallelism involves splitting the input data across GPUs, while model parallelism divides the model across devices. These approaches require efficient data movement and communication to minimize delays during training. 

As such, distributed training frameworks rely on efficient parallelism strategies to handle the growing complexity of large-scale models. 
Distributing either model or data across devices incurs communication, and synchronization methods, such as allreduce using specialized libraries such as NCCL for NVIDIA GPUs and RCCL for AMD GPUs, are commonly used to synchronize gradients or combine activations to ensure efficient scaling of workloads across GPUs~\cite{nccl, rccl}. 
Two widely used paradigms in distributed training, Fully Sharded Data-Level Parallelism (FSDP) and Pipeline Parallelism, adopt techniques to maximize hardware utilization while addressing challenges related to overlapping communication and computation.
Additionally, asynchronous communication features embedded in modern GPU libraries allow for overlapping computation and communication, helping reduce idle times and maximize resource utilization~\cite{pytorch,cudnn,cutlass}.

\niparagraph{Fully Sharded Data-Level Parallelism.}
Data-level parallelism distributes the training data across GPUs, with each GPU computing gradients for a subset of the data. Fully Sharded Data-Level Parallelism (FSDP), implemented in optimizers like ZeRO~\cite{zero, deepspeed}, extends this by partitioning model states, including parameters, gradients, and optimizer states, across GPUs, as shown in Figure~\ref{fig:distributed_training}(a). This partitioning significantly reduces redundant memory usage, allowing larger models to fit into the memory constraints of existing hardware. By enabling distributed training of models that would otherwise exceed hardware capacity, FSDP achieves improved scalability and efficiency. 
One of the key features of FSDP is its ability to overlap computation and communication during the backward pass. For instance, as gradients for a layer are computed, they are asynchronously communicated to the other GPUs. This overlap reduces idle times and ensures that GPUs are fully utilized throughout the training process. However, as model sizes grow, the communication overhead in synchronizing gradients increases, introducing a bottleneck. Techniques such as gradient accumulation and all-to-all reduction optimizations are employed to mitigate these issues, maintaining the efficiency of FSDP~\cite{megatronlm, sparsetransformer}.

\niparagraph{Pipeline Parallelism.}
Pipeline parallelism partitions a model into distinct stages, distributing these stages across GPUs as shown in Figure~\ref{fig:distributed_training}(b). During training, mini-batches flow through the pipeline, enabling different GPUs to concurrently process different batches. Frameworks like PipeDream ~\cite{pipedream, pipedream_flush} and GPipe \cite{gpipe} have demonstrated that efficient pipeline scheduling can overlap forward and backward computations, maximizing throughput. 
A crucial component of pipeline parallelism is its reliance on effective workload balancing. Uneven stage workloads can lead to GPU idling, reducing overall efficiency. Advanced scheduling algorithms aim to minimize idle times by redistributing workloads and introducing interleaved execution patterns. 
For example, PipeDream introduces flush mechanisms to reduce pipeline stalls, while newer approaches integrate communication overlap with computation at a finer granularity to further enhance throughput~\cite{pipedream, pipedream2bw}. Additionally, overlapping the communication of activations and gradients with computation has been shown to reduce overhead, especially in models with deep pipelines~\cite{gpipe}.

\section{Compute and Communication Overlap in Distributed Training}

Overlapping communication and compute is critical for both FSDP and pipeline parallelism. 
By integrating overlapping techniques into these parallelism strategies, distributed training frameworks can fully leverage the capabilities of modern hardware, addressing scalability and performance challenges posed by large-scale models.
Techniques to achieve this include asynchronous gradient communication in FSDP and concurrent forward and backward passes in pipeline parallelism. However, the effectiveness of these overlaps is highly dependent on the communication infrastructure and model characteristics.
As model size grows, the communication latency increases, often negating the benefits of overlap. Strategies such as optimizing all-to-all and reduce-scatter operations have emerged as practical solutions to reduce communication delays. For instance, frameworks like NCCL leverage advanced hardware interconnects, such as NVLink and InfiniBand, to accelerate communication across GPUs~\cite{nccl, rccl}.

Figure~\ref{fig:distributed_training} highlights grey regions representing periods where computation overlaps with communication. In FSDP (Figure~\ref{fig:distributed_training}(a)), model parameters, gradients, and optimizer states are asynchronously communicated to other GPUs during the forward and backward passes, allowing subsequent computations to proceed without delays. Similarly, pipeline parallelism (Figure~\ref{fig:distributed_training}(b)) overlaps the communication of activations and gradients with computations across different stages, effectively maximizing hardware utilization and minimizing idle times.

However, as discussed in the following section, overlapping tasks significantly increase simultaneous resource usage, leading to higher peak power consumption and greater contention for shared resources like memory bandwidth. This contention causes computation slowdowns, reducing the benefits of overlap in certain configurations.

%% file: body/methodology.tex
\section{Methodology}

\input{body/table/gpus}
\input{body/table/models}

\subsection{Evaluated Hardware}
This study examines state-of-the-art hardware accelerators including Nvidia H100 and A100, and AMD MI250 and MI210, as outlined in Table \ref{tab:gpus}. Nvidia GPUs are interconnected using NVLink, with a bandwidth of 900GB/s for H100 and 600GB/s for A100, facilitated by NVSwitch to guarantee peak bandwidth on peer-to-peer connections. AMD GPUs are interconnected using Infinity Fabric with a bandwidth of 300GB/s, which ensures high-bandwidth, low-latency communication for distributed training. These interconnect technologies were critical for the analysis of computation and communication overlap in distributed training. In this paper, we focus on single-node training, where GPUs are located within the same node without internode connections, to isolate and evaluate hardware-specific performance characteristics.

\subsection{Software Framework}
The experiments use PyTorch as the primary training framework, integrated with distributed frameworks such as Megatron-LM for pipeline parallelism and DeepSpeed for FSDP~\cite{megatronlm, deepspeed}. These tools provide optimized implementations for both NVIDIA and AMD hardware, enabling effective resource utilization while minimizing communication bottlenecks. Communication-computation overlap is supported by these frameworks, leveraging the latest versions of PyTorch and hardware-specific libraries to reduce idle GPU time. For NVIDIA GPUs, PyTorch 2.4 compiled with CUDA 12.4 and the NCCL library for collective communications was used, and for AMD GPUs, PyTorch 2.4 compiled with ROCm 6.2 and the RCCL library was used. Differences in communication-computation overlap support were observed between NVIDIA and AMD frameworks, which were attributed to architectural distinctions and further analyzed in the experiments.

\subsection{Workloads}
The workloads included training GPT-3 and LLaMA models, with parameter sizes from 1.3B to 13B. These models were chosen due to their computational demands and relevance to real-world AI applications, providing a representative spectrum of large-scale deep learning tasks. The workloads tested a variety of configurations, as detailed in Table \ref{tab:workloads}, to evaluate the performance and power dynamics of distributed training setups under realistic conditions.

\subsection{Metrics}

Performance metrics included compute kernel time, communication kernel time, and the slowdown of compute kernel time due to overlapping execution. The slowdown is calculated as the relative difference between overlapping and when compute was happening in isolation, defined as:

\footnotesize
\begin{multline}
Compute\ Slowdown \\ 
    = \frac{Compute_{Overlapping} - Compute_{Sequential}}{Compute_{Sequential}},
\end{multline}
\normalsize 

where $Compute_{Overlapping}$ represents sum of compute kernel times during overlapping execution, and $Compute_{Sequential}$ is the sum of compute kernel times when only compute was executed in isolation. 
Additionally, the ratio of overlapped computation was calculated as:

\footnotesize
\begin{multline}
Overlapped\ Computation \\ 
    = \frac{Compute Time_{Overlapped\ with\ Comm}}{Compute Time_{Overlap}},
\end{multline}
\normalsize 

For end-to-end training performance evaluation, we use three metrics, $E2E_{Sequential}$, $E2E_{Overlapping}$, and $E2E_{Ideal}$. 
These represent the latency of a single training iteration for sequential , overlapping, and ideal execution, respectively. 
They are measured (E2E$_{Overlapping}, $Compute$_{Overlapping}$, Compute$_{Sequential}$) and subsequently computed (E2E$_{Ideal}$) as:

\footnotesize
\begin{equation}
\text{Slowdown}_{Compute} = \text{Compute}_{Overlapping} - \text{Compute}_{Sequential}
\end{equation}
\begin{equation}\label{eq:idealtime_e2e}
E2E_{Ideal} = E2E_{Overlapping} - \text{Slowdown}_{Compute}
\end{equation}
\begin{equation}\label{eq:sequential_e2e}
E2E_{Sequential} = E2E_{Ideal} + \text{Overlapped Communication}
\end{equation}
\normalsize

$E2E_{Ideal}$ is a hypothetical time, in which we do not observe the slowdown due to any resource contention, but compute and communication can happen concurrently. 
The ideal execution time, $E2E_{Ideal}$, is computed by subtracting the slowdown in compute time from the overlapping execution time, as shown in Equation~\ref{eq:idealtime_e2e}. 
Sequential execution time, $E2E_{Sequential}$, is measured when all execution including compute and communication happen one after the other.
It essentially also includes the hidden communication kernel time during overlap to the ideal execution time, as defined in Equation~\ref{eq:sequential_e2e}.
We use PyTorch profiler and \texttt{torch.cuda.event} API to measure the training iteration time and execution time of computation and communication kernels.
We also measure average and peak power consumption. NVML was used on NVIDIA GPUs to measure average power consumption over 100ms intervals, while AMD-SMI was utilize on AMD GPUs with a 20ms sampling interval.
NVIDIA-SMI was used for setting power caps during ablation studies. All metrics were averaged over 25 runs to ensure consistency and reliability.

\subsection{Experiment Design}

The experiments systematically analyzed the effects of overlapping computation and communication across various configurations. 
We first measure the computation slowdown, overlapping ratio, end-to-end training iteration time, and power consumption for FP16 training to identify trends across hyper-parameters such as batch size, model size, and parallelism strategy. 
These initial observations, discussed in the following section, provides insights into the interplay between performance and energy efficiency.

Based on these observations, we conduct controlled experiments to characterize the factors contributing to slowdowns and power consumption.
These experiments vary key training configurations, including power caps, numeric precision, and datapath types, to isolate their effects on overlapping execution. 
They enable a detailed analysis of resource contention and trade-offs in distributed training scenarios.

%% file: body/table/gpus.tex
\begin{footnotesize}
\newcommand\ExtraSep
{\dimexpr\cmidrulewidth+\aboverulesep+\belowrulesep\relax}

\newcolumntype{?}{!{\vrule width 2pt}}
\setlength\extrarowheight{3pt}

\begin{table}
\centering
\caption{List of GPUs evaluated}
\resizebox{1.0\columnwidth}{!}
{
\begin{tabular}{ c | c | c | c | c | c}

 \hline
\textbf{Vendor} & \textbf{GPU} & \textbf{Year} & \textbf{Peak FLOPS} & \textbf{Peak FLOPS} & \textbf{Memory Size}  \\
\textbf{} & \textbf{} & \textbf{} & \textbf{(FP32)} & \textbf{(FP16)} & \textbf{(GB)}  \\
 \hline
NVIDIA 
    & A100  & 2020 & 19.5 & 312  & 40  \\
& H100   & 2022 & 66.9 & 1979  & 80  \\
\hline 
\hline
AMD
    & MI210  & 2021 & 22.6 & 181.0 & 64   \\
    & MI250  & 2021 & 45.3 & 362.1 & 128   \\
\hline

\end{tabular}
}

\label{tab:gpus}

\end{table}
\end{footnotesize}

%% file: body/table/models.tex
\begin{footnotesize}
\newcommand\ExtraSep
{\dimexpr\cmidrulewidth+\aboverulesep+\belowrulesep\relax}
\newcolumntype{?}{!{\vrule width 2pt}}
\setlength\extrarowheight{3pt}
\begin{table}
\centering
\caption{Workloads evaluated, detailing model complexity through parameter size and architectural configurations.}
\resizebox{1.0\columnwidth}{!}
{
\vspace{-1.5ex}
\begin{tabular}{ l |  c | c | c | c }

 \hline
\textbf{Model} & \textbf{Parameters} & \textbf{Layers} & \textbf{Attention Heads} & \textbf{Hidden Dimensions} \\
 \hline
GPT-3 XL     & 1.3B  & 24 & 32 & 2048 \\
GPT-3 2.7B   & 2.7B  & 32 & 32 & 2560 \\
GPT-3 6.7B   & 6.7B  & 32 & 32 & 4096 \\
GPT-3 13B    & 13B   & 40 & 40 & 5120 \\
LLaMA 2 13B  & 13B   & 40 & 40 & 5120 \\
\hline

\end{tabular}
}
\label{tab:workloads}
\end{table}
\end{footnotesize}

%% file: body/evaluation.tex
\begin{figure*}[t!]
\centering
{\includegraphics[width=0.95\textwidth]{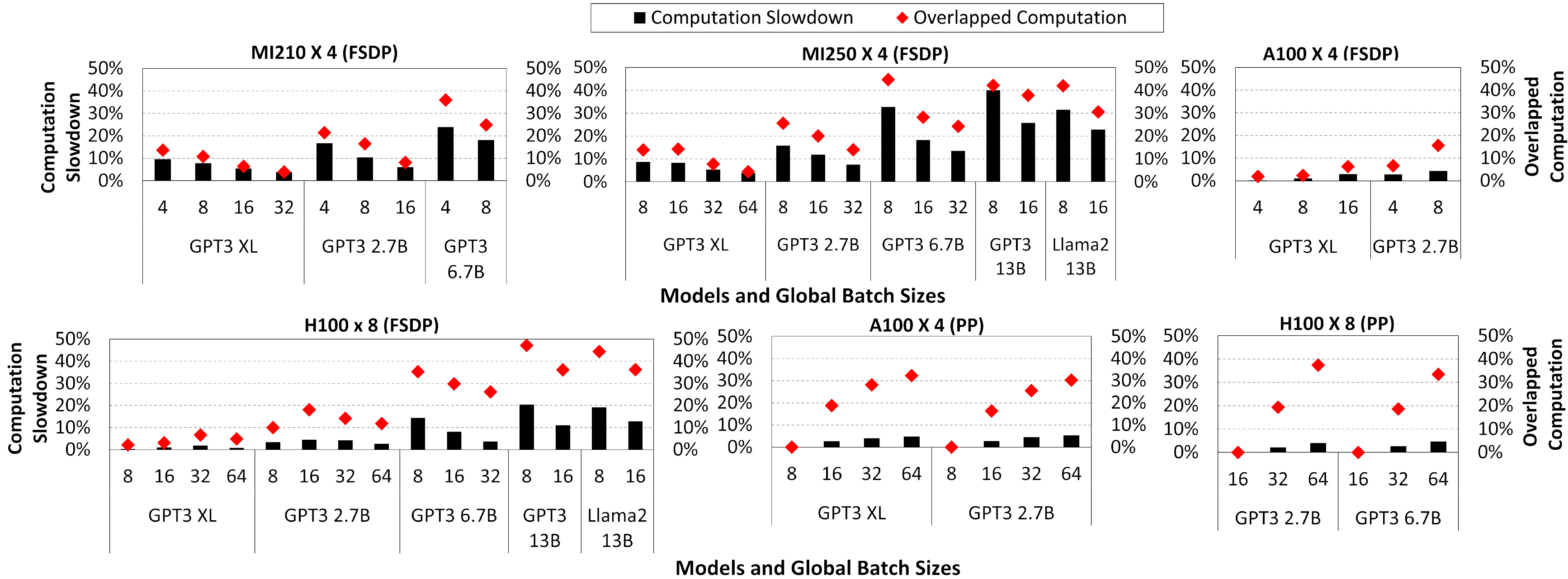}}
\caption{Computation slowdowns across GPUs for various models.}
\label{fig:main_time_slowdown}
\vspace{-1ex}
\end{figure*}

\begin{figure*}[t!]
\centering
\includegraphics[width=0.95\textwidth]{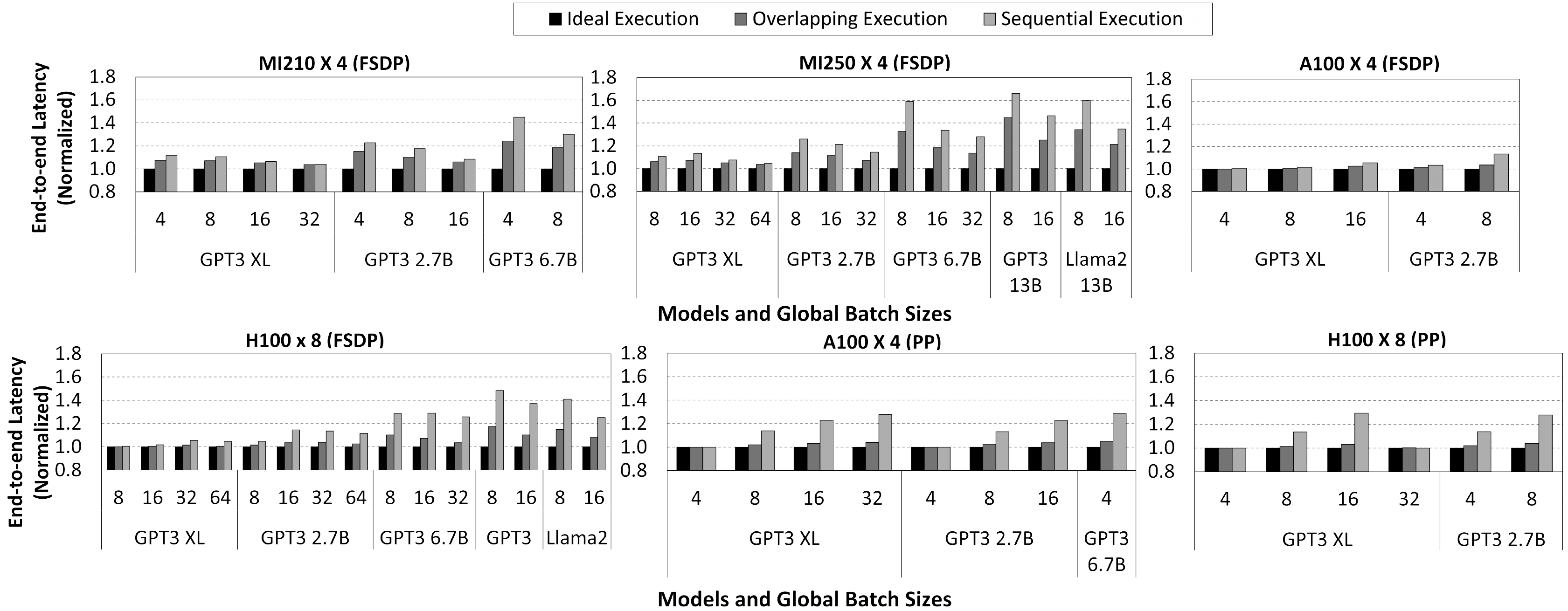}
\caption{End-to-end training iteration latency across GPUs for various models.}
\label{fig:main_time_e2e}
\vspace{-1ex}
\end{figure*}

\section{Evaluation}

This section presents a comprehensive evaluation of overlapping compute and communication across GPUs during distributed training. 
Further, we analyze if power is the constrained resource that slows down compute when executed in tandem with communication. 
Various configurations and training strategies are examined, with a focus on understanding performance and power characteristics of distributed training in context of overlapping compute and communication.

\subsection{Performance Slowdown Analysis}

\niparagraph{Impact of various distribution strategies.}
Figure \ref{fig:main_time_slowdown} illustrates the computation slowdown observed across different GPUs, model sizes, batch sizes, and two parallelization schemes: FSDP and pipeline parallelism.
FSDP partitions model parameters, gradients, and optimizer states across multiple GPUs, necessitating frequent communication to synchronize the partitioned states.
Pipeline parallelism, on the other hand, partitions the model into chunks across GPUs, requiring communication to send and receive intermediate activations.

FSDP observes high overlap ratios, as high as 42\%, but this comes at the cost of performance slowdowns. 
This is because FSDP employs complex communication collectives such as all-gather and reduce scatter, where the latter even requires a compute reduction operation.
For example, the MI210 GPU recorded an average slowdown of 11.3\%, with peak values reaching 23\%. In comparison, the NVIDIA H100 exhibited a broader range of slowdowns from 2.3\% to 7.25\%, peaking at 19.2\%. These slowdowns were predominantly due to resource contention between overlapping computation and communication tasks, which was exacerbated in scenarios involving high overlap ratios.

The NVIDIA A100 showed lower overall slowdowns, with a maximum of 4.3\%. This behavior can be attributed to the A100's smaller memory capacity of 40 GB, which limited its ability to handle larger models. For example, the A100 was constrained to models up to GPT-3 2.7B, with smaller overlap ratios and less contention, resulting in reduced slowdowns.

Pipeline parallelism exhibited different trends compared to FSDP. While slowdowns increased with higher overlap ratios, they consistently remained lower than those observed in FSDP. For instance, on the A100 GPU, from no slowdown for GPT-3 XL with a batch size of 8 to a slowdown of 5.4\% for GPT-3 2.7B with a batch size of 64. This reduced contention stems from pipeline parallelism's less intensive communication patterns, such as send and receive, which mitigate resource contention and result in lower slowdowns under significant overlap scenarios.

\noindent \colorbox{lightgray!50}{\textbf{Takeaway 1:}} \textit{Distribution strategies that rely on complex communication collectives require greater overlap to hide communication latency but, in turn, exhibit higher slowdowns.}

\niparagraph{Impact of batch and model size.}
The relationship between batch size, model size, and performance slowdowns varied significantly across different distribution strategies. In FSDP, larger batch sizes resulted in reduced slowdowns due to the diminished overlapping computation region. This occurred because scaling computation with larger batches outpaced the corresponding scaling of communication. Conversely, pipeline parallelism displayed the opposite trend: larger batch sizes increased slowdowns as the overlapping computation region expanded proportionally. This divergence underscores the necessity for workload-specific optimization strategies tailored to each training paradigm.
Model size also played a critical role in determining performance slowdowns. Smaller models, such as GPT-3 XL, experienced slowdowns of up to 5\% on the H100 using FSDP, while larger models, such as GPT-3 13B and LLaMA 13B, faced slowdowns nearing 40\%. 

\noindent \colorbox{lightgray!50}{\textbf{Takeaway 2:}} \textit{Larger memory footprint and increasing model complexity can have compounding effects  on resource contention and performance degradation due to overlap.}

\niparagraph{Impact of compute and communication overlap on end to end execution.}
Although overlapping computation and communication introduces slowdowns in computation kernels, it provides an effective way to hide the communication overhead introduced by distributed training, thereby improving overall training performance. However, the improved training latency still falls short of the ideal scenario, where no resource contention and thus no slowdown is expected.
Figure \ref{fig:main_time_e2e} compares the execution times of three scenarios: ideal execution without any slowdown, overlapping execution, and sequential execution. 
Our observations show that overlapping execution consistently outperforms sequential execution across GPUs and models. However, it still falls short of ideal execution due to computation kernel slowdowns. For instance, on the MI250 GPU with GPT-3 13B and a batch size of 8, where a 40\% slowdown in computation kernels was observed, the overlapping execution exhibited a 45\% higher execution time compared to the ideal execution time.

\noindent \colorbox{lightgray!50}{\textbf{Takeaway 3:}} \textit{Overlapping computation effectively hides communication overhead and enhances training throughput, but slowdowns in computation kernels limit its full potential.}

\begin{figure*}[t!]
\centering
\includegraphics[width=1\textwidth]{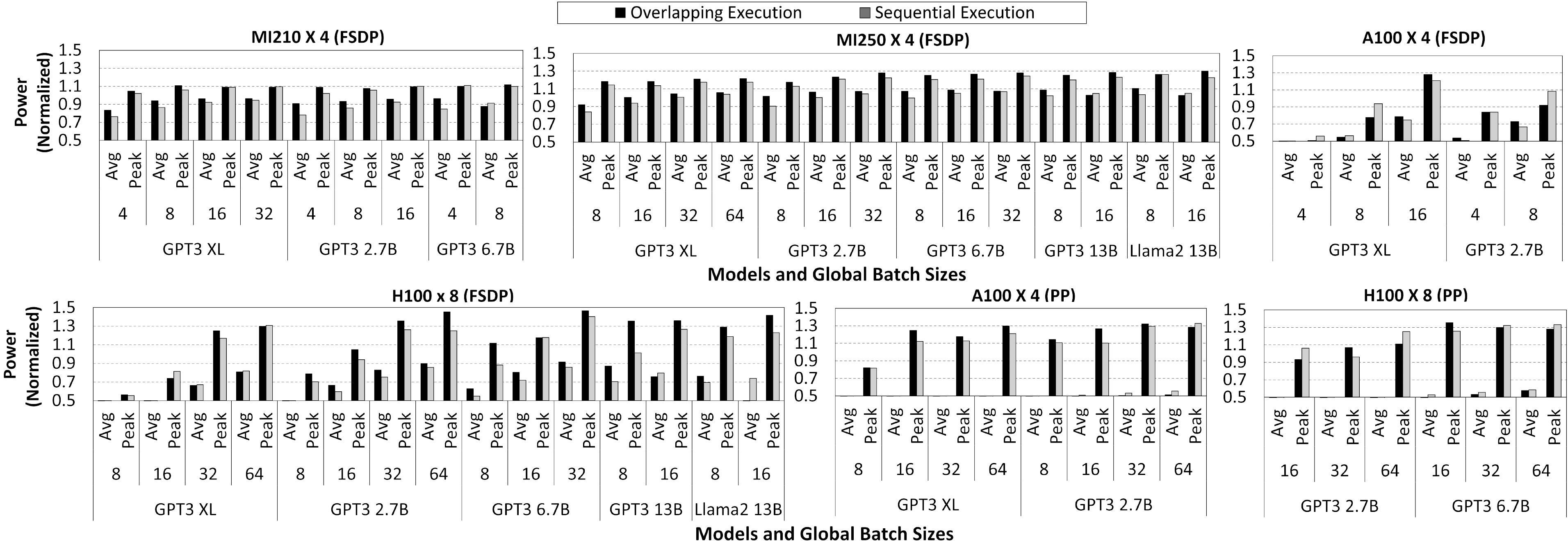}
\caption{Power consumption across GPUs for various models.}
\label{fig:main_power}
\vspace{-1ex}
\end{figure*}

\begin{figure*}[t]
\centering
\includegraphics[width=0.8\textwidth]{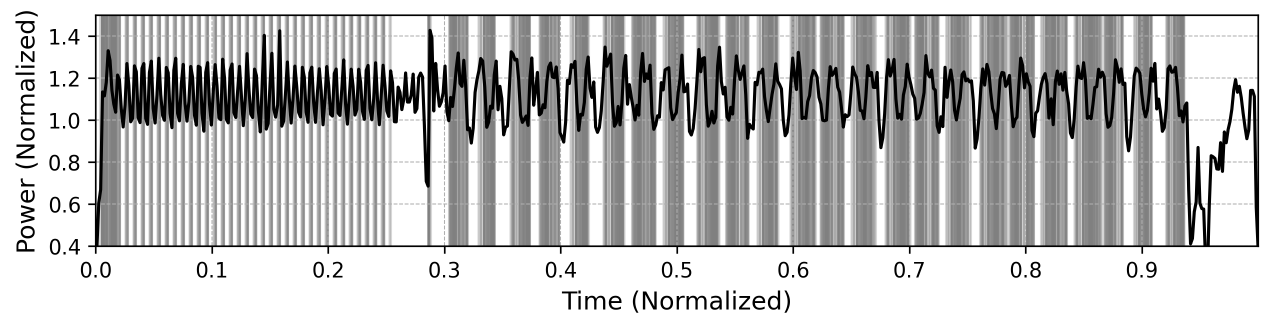}
\caption{Power trace of MI250 during LLaMA2 13B training. Power is normalized to the TDP, and time is normalized to the single iteration (forward and backward pass). The periods of overlapping computation and communication are highlighted.}
\label{fig:power_trace}
\vspace{-1ex}
\end{figure*}

\begin{figure*}[t]
\centering
\includegraphics[width=0.9\textwidth]{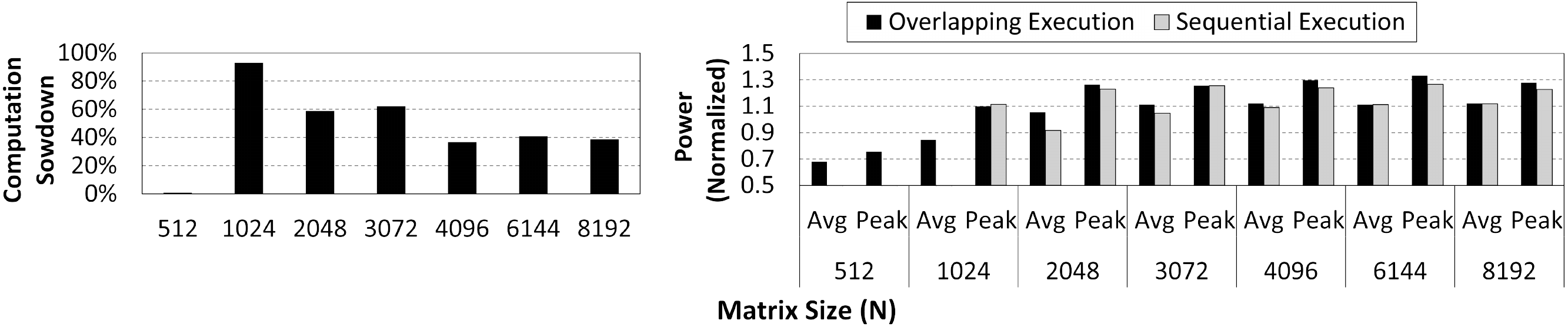}
\caption{Power consumption analysis and performance variations with different overlap configurations. A 1GB all-reduce operation is executed concurrently with matrix multiplication of size NxN.}
\vspace{-2ex}
\label{fig:microbenchmark}
\end{figure*}

\subsection{Power Consumption Analysis}

Power is one of the constrained resource on a GPU, with TDP increasing with every new generation due to higher capabilities and complex architectures.
In this section, we analyze the power consumption of the GPU in this overlapped compute and communication scenario. 
As illustrated in Figure \ref{fig:main_power}, provides additional insights into the energy implications of overlapping computation and communication. 
The H100 GPU consumed an average of 38\% of its TDP during smaller workloads but peaked at 140\% for larger models. These increasing power peaks coincided with greater slowdowns, emphasizing the close relationship between energy consumption and performance bottlenecks. Notably, overlapping scenarios exhibited up to 25\% higher peak power consumption compared to non-overlapping cases.

\noindent \colorbox{lightgray!50}{\textbf{Takeaway 4:}} \textit{Overlapping computation and communication increases peak power consumption, intensifying resource contention and exacerbating performance slowdowns.}

\begin{figure*}[t]
\centering
\includegraphics[width=0.9\textwidth]{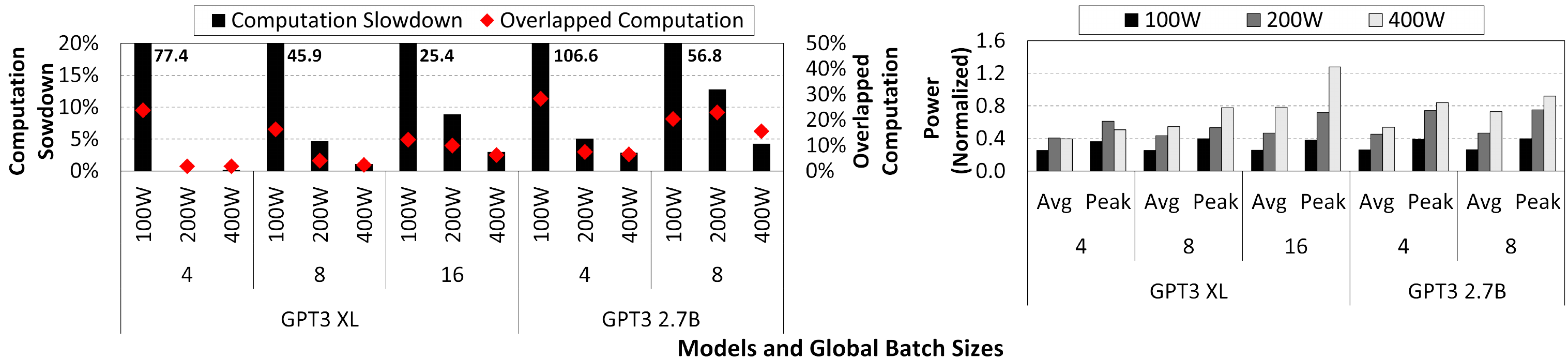}
\caption{Impact of power capping on performance and slowdowns. Evaluated on A100 x 4 systems.}
\vspace{-1ex}
\label{fig:power_capping}
\end{figure*}

\begin{figure*}[t!]
\centering
\subfloat[Computational Slowdown]{\includegraphics[width=1.0\textwidth]{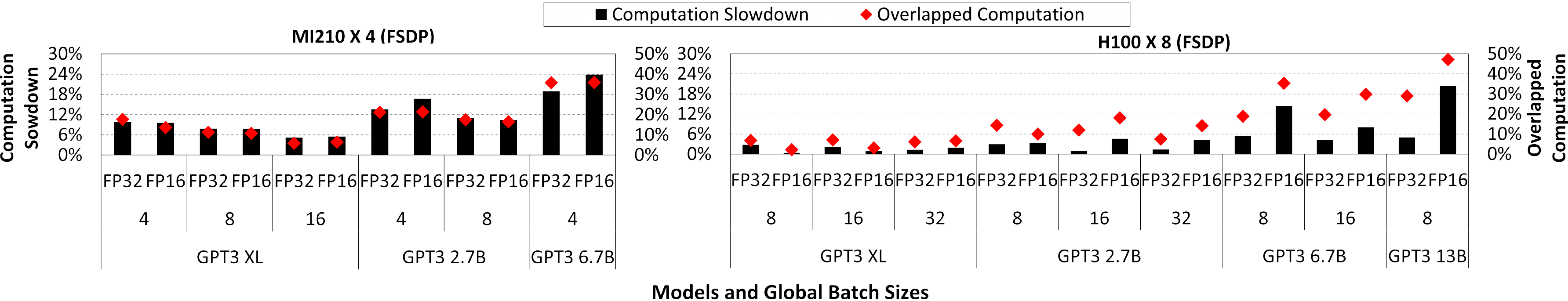}}\\
\subfloat[Power]{\includegraphics[width=0.95\textwidth]{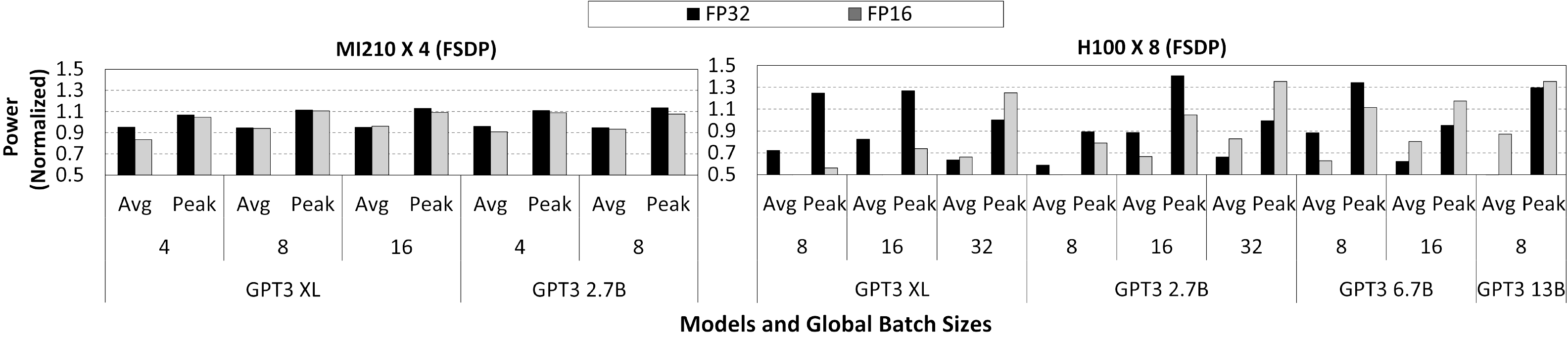}}
\caption{The effect of numeric precision (FP32 and FP16) on slowdowns and power consumption across different workloads.}
\label{fig:numeric_eval}
\vspace{-2ex}
\end{figure*}

\niparagraph{Time series analysis of power usage.}
The time trace presented in Figure \ref{fig:power_trace} highlights power spikes during LLaMA2 13B model training on 4 x MI250 systems. 
We analyze this for AMD systems as ROCM SMI offers a finer granularity power measurement of as low as 1 ms, whereas Nvidia SMI only offers granularity of 100 ms.
These spikes correspond to periods of overlapping computation and communication, as indicated by the gray regions. 
Around time 0.3, we observe highest power spikes due to overlapping compute and communication. 
This observation supports the findings in Figure \ref{fig:main_power}, where overlapping execution increases power consumption.

This observation aligns with findings from the power capping study shown in Figure \ref{fig:power_capping}. Here, experiments conducted on a 4 x A100 system with varying power limits revealed that decreasing power caps exacerbated slowdowns, indicating power resource contention as a factor. 
For example, under a stringent 100W power cap, overlapping execution resulted in slowdowns of up to 100\%, emphasizing the power contention between computation and communication.

\noindent \colorbox{lightgray!50}{\textbf{Takeaway 5:}} \textit{Power constraints contribute to resource contention, leading to slowdowns in overlapping execution.}

\niparagraph{Power analysis on microbenchmarks.}
This experiment, illustrated in Figure \ref{fig:microbenchmark}, examined the effects of overlapping computation and communication on a microbenchmark comprises matrix multiplication overlapping with all-reduce collective. 
Matrix multiplication is executed concurrently with a 1 GB all-reduce operation, and performance degradation is compared to non-overlapping scenarios without concurrent communication.
The results showed that overlapping execution increased both average and peak power consumption, intensifying resource contention and causing greater slowdowns. GPUs often operated near or beyond their TDP limits under these conditions.
These findings indicate that the increased power consumption during overlapping execution is closely linked to resource contention. Potential mitigation strategies include optimizing workload scheduling and improving the management of overlapping execution.

\noindent \colorbox{lightgray!50}{\textbf{Takeaway 6:}} \textit{Overlapping computation and communication increases power consumption and intensifies resource contention, leading to slowdowns, particularly when GPUs operate near or beyond their TDP limits.}

\subsection{Impact of Numeric Precision and Specialized Cores}

\niparagraph{Numeric precision.}
Numeric precision experiments, shown in Figure \ref{fig:numeric_eval}, demonstrated the impact of using lower-precision formats such as FP16. For smaller models like GPT-3 XL, FP16 effectively reduced peak power usage from 1.2x TDP to 0.5x TDP on the H100. However, for larger models and batch sizes, FP16 led to higher overlap ratios, increasing peak power consumption and slowing down performance. This is due to the higher efficiency of the FP16 datapath, which utilizes the hardware more intensively compared to FP32, leading to increased power consumption and, eventually, higher slowdowns.

\begin{figure*}[t!]
\centering
\subfloat[Computational Slowdown\label{fig:tc_slowdown}]{\includegraphics[width=1.0\textwidth]{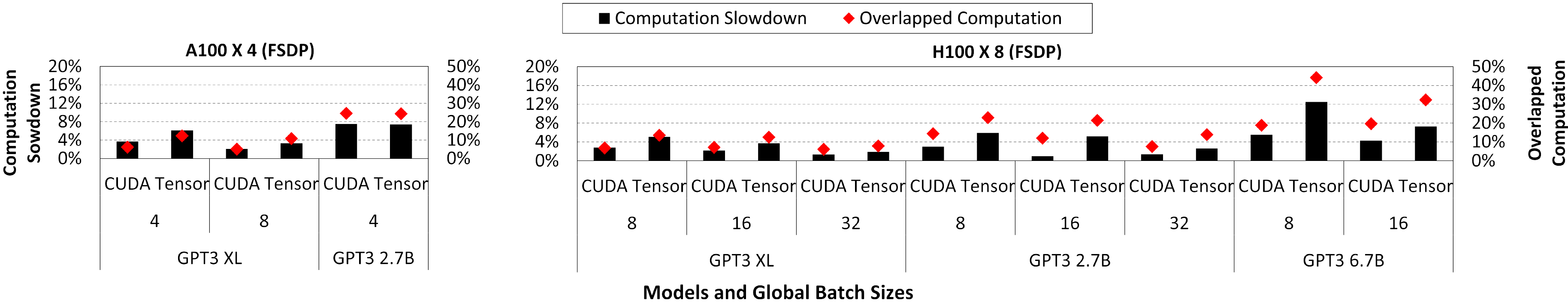}} \\
\subfloat[Power\label{fig:tc_power}]{\includegraphics[width=0.85\textwidth]{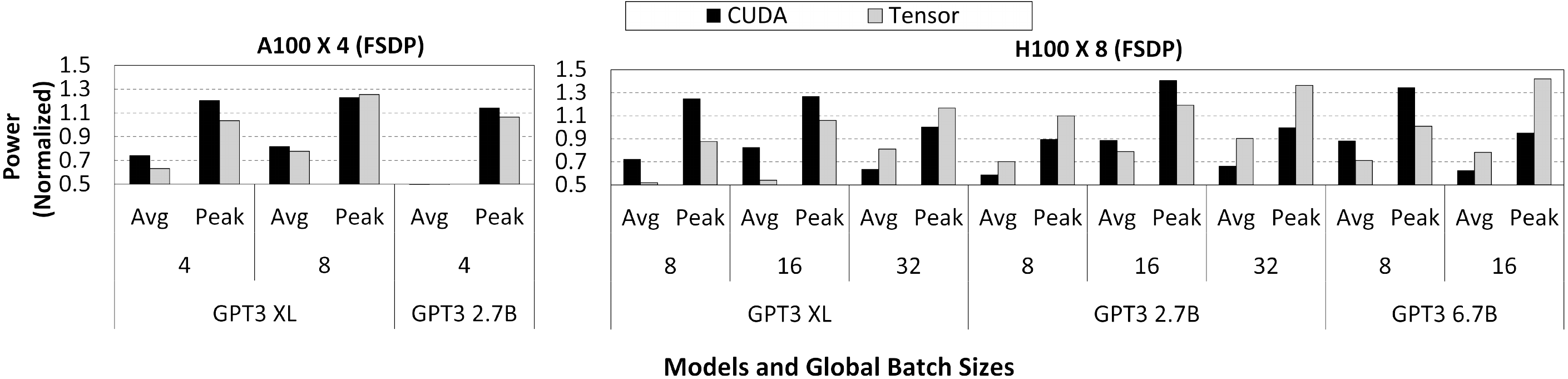}}
\caption{Impact of Tensor Core utilization on performance and power consumption for various workloads.}
\label{fig:tc_eval}
\end{figure*}

\niparagraph{Tensor cores.}
Specialized datapaths, such as Tensor Cores designed for matrix multiplication with lower bitwidth, were analyzed to understand their effects on overlapping computation-communication interactions, as shown in Figure \ref{fig:tc_eval}. 
Tensor cores are specifically designed to efficiently execute certain kernels such as GEMMs and convolutions~\cite{dnnweaver, eyerissv2, tpuv4_isca}.
The comparison involved using FP32 on general compute paths versus Tensor Core execution by converting FP32 to TF32. 
Using Tensor Core demonstrated similar patterns to FP16 precision, achieving reduced average and peak power consumption for smaller models and workloads.
For instance, Tensor Core usage for GPT-3 XL reduced peak power consumption from 1.2$\times$ TDP to 1$\times$ TDP.
However, as model size and batch size increased, resource contention became more pronounced, leading to higher overlap ratios and subsequent slowdowns. 
For example, GPT-3 6.7B with a batch size of 16 exhibited an increase in slowdowns from 4.3\% to 7.3\%, alongside a rise in peak power consumption from 0.95$\times$ TDP to 1.42$\times$ TDP. 
This behavior indicates that while Tensor Cores enhance computational efficiency, they also exacerbate the impact of overlapping tasks on power consumption and resource contention, particularly under high-demand scenarios.

\noindent \colorbox{lightgray!50}{\textbf{Takeaway 7:}} \textit{While lower precision and specialized datapaths enhance efficiency, they also intensify resource contention for larger workloads.}

%% file: body/related.tex
\section{Related Work}

\niparagraph{Optimizing Distributed Training.}
As models continue to grow in size, distributing them has become essential. 
Recent distribution strategies address communication overhead by employing aggressive and opportunistic computation-communication overlap.
Recent works~\cite{zero, deepspeed, fsdp, pipedream, lancet, domino, tutel, lina} on optimizing distributed training overlap communication collectives such as all-reduce, all-gather, reduce-scatter, and send/receive patterns with compute. 
DeepSpeed's Zero Op~\cite{zero}, PyTorch FSDP~\cite{fsdp}, and PipeDream~\cite{pipedream} employ overlapping strategies during FSDP and pipeline parallelism training. 
Domino~\cite{domino} introduces tensor slicing and partitioning techniques to create independent computation and communication chunks for overlapping. 
Lancet~\cite{lancet}, Tutel~\cite{tutel}, and Lina~\cite{lina} optimize mixture-of-experts models by focusing on all-to-all communication required for exchanging expert activations across devices. 

These methods partition expert activation operators and overlap tasks to reduce bottlenecks, often assuming that maximizing overlap is inherently beneficial. While overlapping computation and communication can help mitigate the communication overheads of distribution, these approaches typically overlook resource contention and power analysis, leaving the broader implications of overlap unexplored.

\niparagraph{Mixed Precision and Specialized Hardware.} Mixed precision training, supported by accelerators~\cite{maeri, tabla, dnnweaver, cosmic, dana} has been included in GPUs with NVIDIA Tensor Cores and AMD Matrix Cores, has become integral to enhancing computational efficiency without compromising model accuracy. Reduced precision formats like FP16 and BF16~\cite{mixed,amp} accelerate training while reducing power consumption. However, these techniques require methods such as loss scaling to maintain numerical stability. Emerging formats like TF32 further balance accuracy and computational efficiency, expanding mixed precision applicability in training scenarios for large-scale AI models.

\niparagraph{Power Efficiency in GPUs.}
Power efficiency is a critical concern in distributed machine learning training as workloads scale across devices.
Several works~\cite{splitwise, polca, poweroversub, energyproportion} address this challenge through dynamic power allocation.
SplitWise~\cite{splitwise} employs adaptive power allocation tailored to the different phases of generative LLM inference, while POLCA~\cite{polca} uses power capping strategies to balance performance and energy consumption.
Other approaches~\cite{zeus, envpipe} focus on optimizing system or hardware parameters to reduce energy costs.
Zeus~\cite{zeus} minimizes the energy consumption of machine learning training by optimizing batch size, while EvnPipe~\cite{envpipe} enhances energy efficiency in pipeline parallelism by codesigning clock frequency reduction with operator scheduling tailored for pipeline execution.
However, none of these works analyze the slowdown or power implications caused by compute-communication overlap.

%% file: body/conclusion.tex
\section{Conclusion}

This work provides a detailed characterization of GPU-accelerated systems for distributed deep learning, focusing on the performance slowdowns caused by compute-communication overlap.
To better understand resource contention on GPU devices, this study further examines power consumption when computation and communication occur simultaneously.
Evaluations on NVIDIA and AMD GPUs showed that while overlapping strategies mitigate some of the communication overheads due to distribution, their benefits are constrained by resource contention, causing higher power consumption and performance slowdowns, especially with larger models and batch sizes.
Ablation studies show that lower precision formats and specialized datapaths such as tensor cores improve efficiency for smaller models but intensify resource contention in larger workloads. Power and frequency capping effectively reduce energy consumption but incur performance trade-offs under strict limits.
These findings highlight the need for distributed training frameworks to carefully balance overlapping strategies and resource contention to achieve both performance and power efficiency.

%% file: body/ack.tex
\section{Acknowledgements}

This research was supported through computational resources provided by Partnership for an Advanced Computing Environment (PACE) at Georgia Tech, Google Cloud, and AMD AI \& HPC Cluster.
This work is partially supported by Gifts from Google and AMD. 
We thank Srilatha Manne, Zicheng Liu, and Nathaniel Morris for their valuable feedback and in-depth technical discussions. The paper would not be possible without their input and support.  
The views and conclusions contained herein are those of the authors. They should not be interpreted as representing the official policies or endorsements, either expressed or implied, of Georgia Tech or AMD.